\documentclass[10pt, a4paper]{article}

\usepackage{styles/arxiv}         
\usepackage[utf8]{inputenc}

\usepackage[english]{babel}
\usepackage[T1]{fontenc}    
\usepackage{hyperref}       
\usepackage{url}            
\usepackage{booktabs}       
\usepackage{amsfonts}       
\usepackage{microtype}      
\usepackage{amsmath}
\usepackage{amssymb}        
\usepackage{amsthm}         
\usepackage{dsfont}         
\usepackage{float}
\usepackage{graphicx}
\usepackage{stmaryrd}       
\usepackage{tikz}
\usepackage{multirow}       
\usepackage{worldflags}     

\usepackage{pgfplots}
\usepgfplotslibrary{colormaps}

\usepackage{tabularx}       
\usepackage{csquotes}
\usepackage{todonotes}
\usepackage{soul}           
\usepackage{marvosym}       
\usepackage{caption}
\usepackage{subcaption}

\usepackage{listings}       

\usetikzlibrary{positioning, matrix}

\tikzset{%
  every neuron/.style={
    circle,
    draw,
    minimum size=1.5cm
  },
  neuron missing/.style={
    draw=none, 
    scale=1,
    text height=0.5cm,
    execute at begin node=\color{black}$\vdots$
  },
}
\definecolor{codegray}{rgb}{0.5,0.5,0.5}
\definecolor{backcolour}{rgb}{0.95,0.95,0.92}

\lstdefinestyle{mystyle}{
    backgroundcolor=\color{backcolour},
    commentstyle=\color{blue},
    keywordstyle=\color{black},
    stringstyle=\color{black},
    basicstyle=\ttfamily\tiny,
    breaklines=true,
    escapeinside={(*}{*)}
}
\newcommand\setrow[1]{\gdef\rowmac{#1}#1\ignorespaces}
\newcommand\clearrow{\global\let\rowmac\relax}
\clearrow

\usepackage[citestyle=nature]{biblatex}
\pgfplotsset{compat=1.18}
\pgfplotsset{
    colormap={uni}{rgb255(0cm)=(0, 170, 220); rgb255(1cm)=(255, 20, 11)}
}
\pgfplotsset{
    colormap={greenred}{rgb255(0cm)=(255, 20, 11); rgb255(0.1cm)=(204, 0, 0); rgb255(0.25cm)=(0, 0, 0); rgb255(0.4cm)=(51, 102, 0); rgb255(1cm)=(0, 220, 0)}
}

\graphicspath{ {./figures/} }

\title{AI for Handball: predicting and explaining the 2024 Olympic Games tournament with Deep Learning and Large Language Models}

\date{} 					

\author{
 Florian Felice\footnote{Work} \\
 Department of Mathematics\\
 University of Luxembourg\\
 \href{mailto:florian.felice@uni.lu}{\texttt{florian.felice@uni.lu}}\\
}

\begin{document}

\maketitle

\begin{abstract}
    Over summer 2024, the world will be looking at Paris to encourage their favorite athletes win the Olympic gold medal.
    In handball, few nations will fight hard to win the precious metal with speculations predicting the victory for France or Denmark for men and France or Norway for women.
    However, there is so far no scientific method proposed to predict the final results of the competition.
    In this work, we leverage a deep learning model to predict the results of the handball tournament of the 2024 Olympic Games.
    This model, coupled with explainable AI (xAI) techniques, allows us to extract insightful information about the main factors influencing the outcome of each match.
    Notably, xAI helps sports experts understand how factors like match information or individual athlete performance contribute to the predictions. Furthermore, we integrate Large Language Models (LLMs) to generate human-friendly explanations that highlight the most important factors impacting the match results.
    By providing human-centric explanations, our approach offers a deeper understanding of the AI predictions, making them more actionable for coaches and analysts.

\end{abstract}

Keywords: Handball $\cdot$ Deep Learning $\cdot$ explainable AI $\cdot$ Large Language Models $\cdot$ Olympic Games $\cdot$ Paris 2024

\section{Introduction}

\subsection{Motivation}

The Olympic Games are one of the main sports events in the world with around 3,000 athletes from 80 countries competing in 40 different sports.
One of the team-based sports is handball, for which the Olympic tournament is considered as the most prestigious competition.

The 2024 edition has a particular flavor for handball because they are hosted in Paris, France where the national teams for both men and women are considered as favorites.
Indeed, both teams won the gold medal in the last 2020 occurrence in Tokyo, Japan.
In this work, we propose a prediction model which leverages statistics, deep learning and large language models (LLM) to predict and explain the results of men and women handball tournaments.
Previous work from \cite{felice_predicting_2023} proposed to compare different approaches and concluded that tree-based models provide the best predictive performance.
We however realize that all features are numeric which can make it challenging to interpret for sports experts.
In this work, we will explore an alternative to better account for the composition of the teams by adding the lineup information.
Furthermore, with the recent raise of artificial intelligence (AI), and LLMs at the forefront, more tools can assist sports experts interact with complex predictive solutions.
We view AI as a new layer to act between our complex set of predictive, explainable solutions and sports experts such as coaches.
By proposing an AI based solution, which better accounts for the composition of the teams and generate human-friendly outcomes and explanations, we aim to make these advanced predictive solutions accurate and actionable.

\subsection{Literature review}

To predict the results of handball matches, few different methods exist.
\cite{karlis_modelling_2024} proposed a univariate approach by predicting the score difference between the two teams.
They leverage statistical techniques (such as the Skellam distribution and copulas) to compensate for the challenge of non equi-dispersed target variables.
Focusing on the prediction of the scores of both teams, \cite{groll_prediction_2019} proposed a lasso regression with a constant and low variance to account for under-dispersion.
With the same objective of predicting the score of the two teams, \cite{felice_predicting_2023} proposed a Machine Learning (ML) approach accounting for different factors that can impact the score.
In particular, the approach uses the estimation of teams' strengths modelled by a Conway-Maxwell-Poisson distribution \cite{felice_ranking_2024}.
In this work, we will extend this work by provided a deep learning based approach to better account for teams' compositions and provide the lineups as a categorical input feature.

To account for player's information, some approaches \cite{wang_player2vec_2024} use Natural Language Processing by leveraging Transformers \cite{vaswani_attention_2023} to model sequences of events in a match.
This approach allows to learn a numerical representation of the players which can be assimilated as playing behavior.

The Transformer architecture has been key in driving the recent revolution of Artificial Intelligence lead by the major advances of LLMs.
Large Language Models have drastically impacted the field of data science and now allow a wider audience to access and interact with AI systems.

Despite their large success in the industry\footnote{Generative AI market size is projected to reach \$356bn in 2023, from \$5.5bn in 2020 \cite{noauthor_generative_nodate}}, Large Language Models are not widely adopted in the field of sports analytics yet.

To date, in the domain of sports, Large Language Models mostly appear for benchmarking existing models on sports tasks \cite{xia_sportqa_2024, hu_sportsmetrics_2024} and evaluate their reasoning and sports knowledge.
Some benchmarks go further and are used to evaluate the capabilities to summarize play-by-play match events \cite{hu_sportsmetrics_2024}.

These models are also known for their summarization capabilities which can be leveraged to automatically summarize information from the news \cite{zhang_benchmarking_2024}.
Building on this strength, \cite{cheng_snil_2024} proposed a workflow to generate automatic journalistic insights from sports news.
Such tool can act as a first step towards closing the gap between advanced analytics and sports fans and experts.

In this work, we leverage three core components of AI to generate handball match predictions with insightful comments.
We use deep neural networks \cite{goodfellow_deep_2016} to predict the score of two teams in a match based on several attributes.
We then use explainability methods to augment the predictions with an understanding of how the different covariates lead to the predicted score.
Finally, to make these predictions and explanations human-friendly, we leverage a Large Language Model 

This document is therefore structured as follows.
Section \ref{sec:material} details the data used and the core methodology composed of the three AI components.
Section \ref{sec:results} presents the final results with the predictions of the 2024 Olympic tournament for men and women.
It also illustrates the AI-generated explanations for an important match: the final of the tournament.
In Section \ref{sec:conclusion}, we discuss the limitations, potential extensions of this work before concluding.


\section{Materials and methods} \label{sec:material}

Our approach aims to predict the number of goals scored by the home (denoted with the subscript $h$) and away (with subscript $a$) team.
We design a Machine Learning model that learns from past matches to understand how the different factors will impact the score of the teams.
In this section, we first present the data used for training our model and detail the different types of features it uses.
Next, we present the learning approach.
It is based on a neural network which uses information from another model trained on clubs' matches as pre-trained information for initialization.
Last, we show how we can combine xAI techniques and LLMs to explain the predictions of our model in a human-friendly way.

\subsection{Data}\label{sec:data}

In our setup, we design a Machine Learning model for a multi-target regression which aims to predict two outcomes $(y_{h}, y_{a}) \in \mathbb{R}^{2}_{+}$.
We denote $y_h$ the number of goals scored by the home team at the end of a match and $y_a$ is the score of the away team.

We build our dataset using the handball API from \href{https://sportdevs.com/handball}{SportDevs}.
The dataset is multi-modal and combines numerical and textual data.
We thus define four different feature types.
The first three feature types for the numerical covariates are: match information, teams information and teams strengths.
The last feature type is textual and contains the teams lineups.

\paragraph{Match information}
These numerical features aim to carry information about the match and its importance.
It can help gather information such as potential stress of players.

\begin{itemize}
    \item \texttt{Day of week}: encoded day of the week for the start time of the match.
    \item \texttt{Hour}: hour of the start time of the match. We can expect that matchs starting early in the day (e.g., morning) can be less important or teams may lack time for preparation.
    \item \texttt{Importance}: carries the importance of the competition from the lowest (friendly games with value 4) to the highest importance (Olympic Games with value 10). Detailed values are presented in Table \ref{tab:importance} in Appendix \ref{apdx:importance}.
\end{itemize}

\paragraph{Teams information}

We incorporate inherent information of the teams regarding their composition, homogeneity and propensity to fatigue. 

\begin{itemize}
    \item \texttt{Travel distance}: distance in kilometers (as the crow flies) to travel for the home and away teams between the teams locations and the match location. This aims to capture the potential fatigue caused by the travelling distance.
    \item \texttt{Number of clubs}: number of clubs in which each player of the team are playing. A low number suggests that majority of players are from the same clubs and are used to playing together.
\end{itemize}

\paragraph{Teams strengths}

We add features that correspond to the teams’ strengths as described in \cite{felice_ranking_2024}.
These covariates in the spirit of Statistically Enhanced Learning (SEL, \cite{felice_statistically_2023}) aim to estimate the offensive and defensive strengths of both teams.

\begin{itemize}
    \item \texttt{Attack strength}: estimated strength in attack via SEL for home and away teams.
    \item \texttt{Defense strength}: estimated strength in defense via SEL for home and away teams.
\end{itemize}

\paragraph{Teams lineups}

The last feature type corresponds the textual covariate to incorporate the composition of the teams.
We note that, depending on the competition, teams can include up to 16 players\footnote{Competitions such as world championships can count up to 16 players per team per match while Olympic games restrict to 14 players} per match.

\begin{itemize}
    \item \texttt{Home lineup}: list of (up to 16) players for the home team present on the match report.
    \item \texttt{Away lineup}: list of (up to 16) players for the away team present on the match report.
\end{itemize}

This feature requires to know the composition of the teams prior to a match.
To be ingested by the model defined in Section \ref{sec:model}, the list of players of the home and away teams are concatenated to produce one vector of up to 32 players.
This vector needs to be pre-processed to generate a numerical vector composed of word tokens.
Our tokenizer simply considers the full name of the player as one token.

The main challenge remains the amount and the quality of data.
The lineups are not systematically recorded and made available in the API, therefore we use imputation tricks to backfill matches with missing lineups.
In particular, for each team, we use the last lineup available to fill the missing lineups of the season.
Furthermore, we note that the data for modelling club matches are filtered such that lineups are not empty.
We can only enforce such a logic for clubs because of the amount of data available (e.g. for men the training data goes from 55,414 observations to 16,450 data points once filtered, meaning that 38,964 have no lineup information).
For national teams, we allow empty lineups meaning that the model will ingest a vector of null values.
The null value is already a token by itself (with value $=0$), so the model will learn that a vector full of 0's is a specific form of a team.

\begin{table}[ht]
  \centering
  \begin{tabular}{ccccc}
    \toprule
    & Category & Data size & Min date & Max date \\
    \midrule
    \multirow{2}{*}{Men} & Clubs & 16,450 & 2015-08-15 & 2024-05-01 \\
    & National teams & 1,846 & 2015-01-02 & 2024-01-22 \\
    \midrule
    \multirow{2}{*}{Women} & Clubs & 8,100 & 2017-08-16 & 2024-04-24 \\
    & National teams & 1,298 & 2015-03-19 & 2023-12-15 \\
    \bottomrule
  \end{tabular}
  \caption{Comparison of training data size and dates for clubs vs national teams for men and women}\label{tab:data_size}
\end{table}

We observe the large difference of data available between national teams and clubs as presented in Table \ref{tab:data_size}.
We have much more data for clubs than for national teams with a scale for 6.1 times more for women to 8.9 times more for men.
This large difference of available training data will impact the prediction performance.
We will thus use transfer learning as a mitigation technique to benefit from the amount of data available for clubs.

\subsection{Neural networks and transfer learning}\label{sec:model}

\paragraph{Embedding Multi-Layered Perceptron}

Even though the deep learning approaches did not exhibit the best performance amongst the compared methods in \cite{felice_predicting_2023}, neural networks still benefit from an interesting capability we want to build upon.
We want to use the capabilities of these models to handle multi-modal data for explainability purposes.
Indeed, in \cite{felice_predicting_2023}, to exploit the information on the teams' compositions, the tree-based solutions need to ingest numerical features.
The generated features thus included information such as the average and standard deviation of age or height of players per position.
However, when it comes to interpreting the predictions, sports experts gave us the feedback that they had hard times finding a sports meaning from these numerical covariates.
They would instead prefer to be able to identify the name of the players to understand their direct impact on the match.

Building on this important feedback, we update the architecture of the neural network model exposed in \cite{felice_predicting_2023} to better incorporate explainable teams' composition.

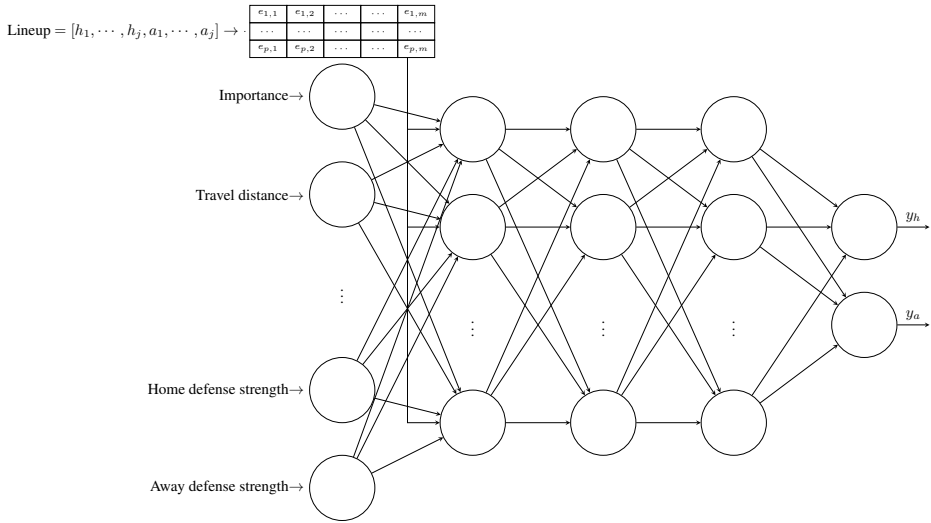
\begin{figure}[ht]
    \centering
    \resizebox{0.75\linewidth}{!}{\begin{tikzpicture}[x=1.5cm, y=1.5cm, >=stealth]

\foreach \m/\l [count=\y] in {1,2,missing,3,4}
  \node [every neuron/.try, neuron \m/.try] (input-\m) at (0,1.5-\y*1.5) {};

\foreach \m [count=\y] in {1,2,missing,3}
  \node [every neuron/.try, neuron \m/.try ] (hidden1-\m) at (2,1-\y*1.5) {};

\foreach \m [count=\y] in {1,2,missing,3}
  \node [every neuron/.try, neuron \m/.try ] (hidden2-\m) at (4,1-\y*1.5) {};

\foreach \m [count=\y] in {1,2,missing,3}
  \node [every neuron/.try, neuron \m/.try ] (hidden3-\m) at (6,1-\y*1.5) {};

\foreach \m [count=\y] in {1,2}
  \node [every neuron/.try, neuron \m/.try ] (output-\m) at (8,-0.5-\y*1.5) {};


\foreach \l [count=\i] in {Importance,Travel distance,Home defense strength,Away defense strength}
  \draw [-] (input-\i) -- ++(-0.5,0)
    node [left, midway, draw=none] {\l $\rightarrow$};

\foreach \l [count=\i] in {h,a}
  \draw [->] (output-\i) -- ++(1,0)
    node [above, midway, draw=none] {$y_\l$};

\foreach \i in {1,...,4}
  \foreach \j in {1,...,3}
    \draw [->] (input-\i) -- (hidden1-\j);

\foreach \i in {1,...,3}
  \foreach \j in {1,...,3}
    \draw [->] (hidden1-\i) -- (hidden2-\j);

\foreach \i in {1,...,3}
  \foreach \j in {1,...,3}
    \draw [->] (hidden2-\i) -- (hidden3-\j);

\foreach \i in {1,...,3}
  \foreach \j in {1,...,2}
    \draw [->] (hidden3-\i) -- (output-\j);


\def\d{0pt}
\matrix (m) at (0,1) [matrix of nodes,
            draw=none,
             nodes={minimum width=0.85cm,
                    minimum height=0.4cm,
                    draw=black, 
                    outer sep=0pt,
                    inner sep=0,
                    line width=\d,
                    font=\tiny,
                    anchor=center},
             nodes in empty cells, column sep=0, row sep=-\d,
              ]
{
    $e_{1,1}$ & $e_{1,2}$ & $\cdots$ & $\cdots$ & $e_{1,m}$  \\
    $\cdots$ & $\cdots$ & $\cdots$ & $\cdots$ & $\cdots$  \\
    $e_{p,1}$ & $e_{p,2}$ & $\cdots$ & $\cdots$ & $e_{p,m}$  \\
};

\foreach \j in {1,...,3}
    \draw [->] (1, 0.6) |- (hidden1-\j);

\draw [-] (-1.5,1) -- ++(0,0)
    node [left, midway, draw=none] {Lineup $= [h_1, \cdots, h_j, a_1, \cdots, a_j] \rightarrow$};

\end{tikzpicture}}
    \caption{Architecture of the multi-modal neural network. The textual inputs are passed to an embedding layer before being merged with the numerical inputs to later be fed to the fully connected part of the network.}
    \label{fig:ann_archi}
\end{figure}

As we introduced in Section \ref{sec:data}, we keep the main features types about match and teams information along with strengths.
To these features, we now add the lineups via an embedding layer as depicted in Figure \ref{fig:ann_archi}.

The embedding layer ingests the concatenated lineup vector $[h_1, \cdots, h_j, a_1, \cdots, a_j] \in \mathbb{N}^{2j}$ and maps the values to a continuous vector of fixed size in $\mathbb{R}^{m}$.
Considering the maximum number of players in different competitions, we set $j=16$.
After tuning hyper-parameters of the model, we set the output dimension $m = 25$, meaning that each player has a vector representation in $m=25$ dimensions.

The weights of this layer are usually initialized with random values pulled from a standard normal distribution $\mathcal{N}(0, 1)$ \cite{ansel_pytorch_2024}.
However, as we will see in the next paragraph, initializing with non-random weights from a pre-trained model via transfer learning can help the training process.

During the training process, the embedding layer learns a numerical representation of each player that corresponds to their intrinsic impact on the match results.
We note that the model does not have information about the player's performance itself (number of goals, etc.).
Instead, it learns how the team performs when the player is present in the lineup.
In other words, it learns a form of plus-minus rating of players \cite{hvattum_comprehensive_2019} but in an implicit manner.

\paragraph{Transfer Learning}

To mitigate the difference in data between clubs and national teams settings, we use a transfer learning approach.
Given that players in national teams also necessarily have records in the clubs data set, we want to leverage the information learnt by the clubs model to support the training of the model for national teams.
Moreover, since we have a larger data set for clubs, and potentially of better quality (since filtered), we expect that the transfer of information will help boost the performance of the model for national teams.

We use the approach called "Network-based deep transfer learning" \cite{tan_survey_2018}.
When initializing the weights of our international model, the weights of the embedding layer are no longer pulled from a normal distribution as the default behavior.
Instead, we set the weights with the values learnt from the pre-trained clubs model.
Since players from national teams were also present in the data when pre-training the clubs model, we expect that their numerical representation will be similar in both models.
The model will then adjust the values during the training process to account for the impact of players in international competitions.

This implies that some non-international players will still have a numerical representation, since available in the pre-training stage, but their weights will no longer be of interest.
We also note that we cannot initialize all the weights of the model from the pre-trained one since the rest of the numerical features are not the same between clubs and international teams models.
For instance, the clubs' model considers the share of international players that will be a constant equal to 100\% for the international model.
On the other hand, the international model considers the number of clubs that we be equal to 1 for the clubs model.

This approach aims to provide some initial knowledge to our model when starting the training so it can benefit from the information learnt by the pre-trained model and later exhibit better performance than with a random initialization.

\subsection{Large Language Model for human readable explanations}

The core objective of this work is not only to provide accurate predictions but also to suggest meaningful explanations of the match.
We play with the gradient-based nature of neural network to use Integrated Gradients to generate local explanations \cite{sundararajan_axiomatic_2017}.
This method is used for attributing a score on the model's input features based on their impact on a prediction.
It works by accumulating the gradients interpolated along a straight-line path from a defined baseline input to the actual input value and integrates the accumulated values.
The resulting attributions provide a measure of the feature's importance that sums to the difference between the model's output for the input and the baseline.

The explanations generated by this method can be hard to read and to interpret for humans.
Additionally, the relative high cardinality the input features\footnote{43 values coming from $2 \times 16 = 32$ players and 11 covariates as described in Section~\ref{sec:data}.} can make this task even more difficult.
To exploit such information, only experts in both handball and explainable Machine Learning would be able to derive meaningful information.
In other words, the raw results without an interpretation layer are of little interest for coaches.
It would require massive educational efforts to give them the required skills.

To mitigate this issue, we use a Large Language Model to extract meaningful information in a short and human-readable summary.
In particular, we use the open-source model Mistral 7B instruct \cite{jiang_mistral_2023}.
The model was trained on a large corpus of text available from the internet corresponding to approximately 600 billion words\footnote{Estimated 800 billion tokens with the rule-of-thumb of 1 token $= \frac{3}{4}$ words.}.
This gives the ability to the model to reason on a large set of topics.
To afford running a LLM on relatively small server (with 8GB of GPU memory), we use a quantized version of the model which allows to compress it and limit its memory footprint \cite{frantar_gptq_2023}.
We leverage the reasoning and summarization capabilities of the model to generate a comprehensive summary of the match based on the predictions and explanations.

To generate the match summary, we structure the prompt in three part: inputs, examples and instructions.
The final prompt template is available in Appendix~\ref{apdx:prompt}.

\paragraph{LLM inputs}

The output of the prediction model and of the xAI function become the input of our LLM.
We first give contextual information to the model about the teams, date of the match, the type of competition and the predicted score.
The LLM knows the expected outcome and can adjust its explanations based on the results and the overall match context.
Then, we provide information about the covariates of the model.
For each of the features, we provide a short description of what it would mean to have a positive attribution score.
This gives more structure to the model and guides it derive meaningful conclusions.
We finally add the generated feature attributions derived from the xAI method that will be the main source of information for the LLM to generate the match summary.
To each player name, we add the position and the team of the player along with the estimated attribution.

\paragraph{Few-shot prompting}

To ensure that the LLM sticks to the desired structure, we provide few examples in the prompt.
This approach, known as \textit{in-context learning} \cite{wang_large_2023}, uses examples in the context (prompt) to help the model learn  the structure of the expected output and what type of reasoning is expected.
We then take few examples of past matches and manually generate reports as we would expect the LLM to generate them.
We ensure that the different examples we generate are consistent in the format and type of conclusions we want to derive so the model does not get confused.
However, we try to make them diversified so it can learn the variety of output we expect and does not repeat non-informative patterns by just mimicking the examples.

\paragraph{Summarization instructions}

The final important step is to provide the exact instructions to the LLM.
While we already provided the different inputs and examples, we need to formalize and write down how we want the model to behave and what we want it to generate.
We instruct the model on how to interact with the provided inputs and how to use the examples.
We also specify what kind of conclusions and reasoning we expect.
This step is crucial to limit the risk of hallucination and guarantee the accuracy of the response \cite{addlesee_grounding_2024}.

\section{Results} \label{sec:results}

To illustrate the performance of our solution, we first evaluate the prediction model on a test set for international competitions and analyze the change in performance using transfer learning.
We next present the expected results for the 2024 Olympic Games tournament using the presented model.
Finally, we present an example of model explanation using the presented Large Language Model.

\subsection{Prediction performance}

First, we notice that the performance presented in Table \ref{tab:perf} shows higher errors than what is presented in \cite{felice_predicting_2023} because we now evaluate the performance for international competitions and not for clubs.
This performance gap is simply explained by the lower volume of data for national teams, an issue for which the transfer learning approach aims to mitigate.

\begin{table}[ht]
  \centering
  \begin{tabular}{cccccc}
    \toprule
    & & \multicolumn{2}{c}{Home} & \multicolumn{2}{c}{Away}\\
    & Transfer learning? & RMSE & MAPE & RMSE & MAPE \\
    \midrule
    \multirow{2}{*}{Men} & No & 5.96 & 16.55\% & 6.15 & 17.02\% \\
    & Yes & 5.63 & 15.17\% & 5.20 & 14.13\% \\
    \midrule
    \multirow{2}{*}{Women} & No & 7.79 & 22.60\% & 7.26 & 21.82\% \\
    & Yes & 6.17 & 17.01\% & 5.87 & 19.27\% \\
    \bottomrule
  \end{tabular}
  \caption{Comparison of the loss (RMSE and MAPE) with and without transfer learning on the test set}\label{tab:perf}
\end{table}

We note from Table~\ref{tab:perf} that transfer learning consistently helps reduce the error of the model on the test set.
This suggests that the information learnt from the embeddings pre-trained on club data benefit the model for national teams.

\begin{figure}[ht]
    \centering
    \begin{tikzpicture}
\begin{axis}[
    x tick label style={draw=none, anchor=north},
    y tick label style={draw=none, anchor=east},
    x tick style={draw=none},
    y tick style={draw=none},
    label style={draw=none},
    axis lines=left,
    xlabel=Epoch,
    ylabel=MSE,
    legend style={draw=none, anchor=north east, font=\tiny},
  ]
    \addplot [no marks, dashed, red] table[x=epoch, y=train_loss_base,col sep=comma]{figures/data/train_history_tl.csv};
    \addlegendentry{Training loss - No transfer learning};
    \addplot [no marks, red] table[x=epoch, y=test_loss_base,col sep=comma]{figures/data/train_history_tl.csv};
    \addlegendentry{Validation loss - No transfer learning};
    \addplot [no marks, dashed, blue] table[x=epoch, y=train_loss_transfer,col sep=comma] {figures/data/train_history_tl.csv};
    \addlegendentry{Train loss - Transfer learning};
    \addplot [no marks, blue] table[x=epoch, y=test_loss_transfer,col sep=comma] {figures/data/train_history_tl.csv};
    \draw[thin, loosely dashed, color=gray] (2,0) -- (2,0.012);
    \node[gray, right] at (2,0.0087) {\tiny Early stopping};
    \addlegendentry{Validation loss - Transfer learning};
\end{axis}
\end{tikzpicture}
    \caption{Evolution of the training and validation loss over epochs with and without transfer learning (on men national teams data). The loss used is the sum of Mean Squared Error for home and away teams.}
    \label{fig:loss_evo}
\end{figure}
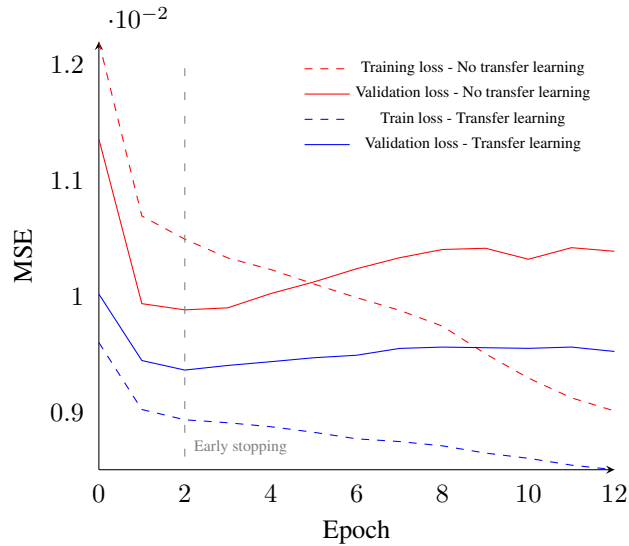

Furthermore, as we can observe from Figure~\ref{fig:loss_evo}, transfer learning greatly benefits the learning process of our model.
Indeed, the curve for the loss (for both train or validation set -- in blue) in the case of transfer learning is always below the one without.
We also observe that over-fitting may occur after several epochs, though the amplitude is much larger in the case without transfer learning.
Indeed, the training loss in both cases continues to decrease while the validation loss plateaus or even increases.
To limit this risk, we apply early stopping to interrupt the training when the validation loss no longer decreases, here after epoch 2 as illustrated in Figure~\ref{fig:loss_evo}.

Interpreting from this plot, we can say that the validation loss evaluated after the first iteration is lower in the case of transfer learning because the weights of the embedding layer are not randomly initialized.
After the second epoch it converges to a minimum value which is always lower than the situation where we do not use transfer learning, leading to better performance.

We can thus observe a transfer of information from the pre-trained clubs model which allows the training of our international model to be smoother and reach a lower level of error.

\subsection{Paris 2024 predictions for handball tournament}

For the 2024 Olympic Games, France is considered as the favorite team for both the men and women handball tournaments.
Furthermore, the last international competitions before the 2024 Olympiads saw France get the gold medal.
Indeed, for the men's European championships in January 2024, France defeated Denmark in final (33 - 31 after extra time) and became European champions.
One month before, France became world champions after winning in front of Norway in the women's world championships (31 - 28).

It is worth mentioning that both French teams also got the gold medal at the 2020 Olympic Games in Tokyo, strengthening the favorite status.

We report the final results from the semi-final and finals using our deep learning solution in Table~\ref{tab:final_rank}.
As expected, we can find the finalists of the last competitions on the podium of the competition.

\begin{table}[ht]
    \begin{subtable}{.5\linewidth}
      \centering
        \begin{tabular}{cl}
            \toprule
            \includegraphics[width=.06\textwidth]{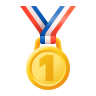} & \worldflag[width=.05\textwidth,length=.07\textwidth]{FR} France \\
            \includegraphics[width=.06\textwidth]{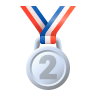} & \worldflag[width=.05\textwidth,length=.07\textwidth]{HR} Croatia \\
            \includegraphics[width=.06\textwidth]{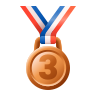} & \worldflag[width=.05\textwidth,length=.07\textwidth]{DK} Denmark \\
            4 & \worldflag[width=.05\textwidth,length=.07\textwidth]{ES} Spain \\
            \bottomrule
        \end{tabular}
        \caption{Men}\label{tab:men_rank}
    \end{subtable}%
    \begin{subtable}{.5\linewidth}
      \centering
        
        \begin{tabular}{cl}
            \toprule
            \includegraphics[width=.06\textwidth]{figures/logos/gold_medal.png} & \worldflag[width=.05\textwidth,length=.07\textwidth]{FR} France \\
            \includegraphics[width=.06\textwidth]{figures/logos/silver_medal.png}  & \worldflag[width=.05\textwidth,length=.07\textwidth]{NO} Norway \\
            \includegraphics[width=.06\textwidth]{figures/logos/bronze_medal.png} & \worldflag[width=.05\textwidth,length=.07\textwidth]{SE} Sweden \\
            4 & \worldflag[width=.05\textwidth,length=.07\textwidth]{DK} Denmark \\
            \bottomrule
        \end{tabular}
        \caption{Women}\label{tab:women_rank}
    \end{subtable} 
    \caption{Top 4 of the final ranking for the handball tournament at the 2024 Olympic Games}
    \label{tab:final_rank}
\end{table}

Analyzing the ranking for the men's tournament in Table~\ref{tab:men_rank}, we expect to see a semi-final between the 2 leading teams France and Denmark, ending in the favor of the local host.
The final will then see France face off against Croatia to end with a new Olympic title for the French team.

On the women's side, as we can observe in Table~\ref{tab:women_rank}, we expect the repetition of the last world championships where France will play against Norway which will also lead to a gold medal for France.
On the bronze final, we expect that Sweden will take their revenge from the bronze final of the 2023 world championships and win against Denmark.

We present more detailed results in Appendix~\ref{apdx:results}.

\subsection{AI-generated explanations}

To go beyond the predictions of the number goals scored by each team, we look at the drivers of the prediction.
For the sake of illustration, we focus on the final match of the men's tournament between France and Croatia.
We expect to see France win the goal medal with a score of 35-24 (see Figure~\ref{fig:final_round_men}).

To have a better understanding about the drivers of the prediction, we first generate the feature attributions using Integrated Gradients from the deep neural network as presented in Section~\ref{sec:model}.
This local explainability method helps us understand how the model uses the different input features to predict the score.
Focusing on one team, say France, we plot the generated attribution values per feature in Appendix~\ref{apdx:local_explain_ig}.
We can see that the relatively large amount of features make the plot interesting only in case we want to dive into more details and potentially compare one feature with another.

To generate a human-friendly summary of the match, we ask the LLM to take the expert translator role and summarize the attribution scores we extracted.
We report the AI-generated output in Listing \ref{list:llm_example}.

\begin{lstlisting}[caption=Output example for the match France vs Croatia]
(*\label{list:llm_example}*)
Report for the upcoming handball match between France and Croatia at the Olympic Games on August 11, 13:30:0

The predicted score is in favor of France with a difference of 11 goals. This suggests a strong attacking capability and solid defense for the French team. Several factors are driving this prediction.

First, the number of clubs represented in the French squad is a positive contributor to the team's performance. With a low number of clubs, French players are more likely to have familiarity and better synergy with each other.

Among the players, the backline is particularly influential. Dika Mem, a right back for France, and Elohim Prandi, a left back, both have significant positive impacts. Their experience and understanding of their positions help solidify the defense, creating opportunities for the team's attack to thrive.

Defensively, both teams show strength. However, France's defense has a slightly more positive impact. This could mean that the French defense is more stable and reliable in comparison to Croatia's.

International experience is also a positive factor for France, as having players familiar with high-pressure competition could translate to better performance on the court.

On the Croatian side, Igor Karacic, a center back, and Luka Lovre Klarica, a right back, have minor negative impacts on the French prediction. This could indicate that these players are effective in their respective roles and could potentially make things difficult for the French attack.

The day of the week and the hour of the match do not play significant roles in the prediction. However, the importance of the competition itself is a strong driver for motivation and intense performance, potentially leading to more goals being scored overall.

Overall, the French team is predicted to have a strong edge in this match due to their collective backline performance, familiarity with each other, and international experience. A solid defense and the ability to capitalize on opportunities will likely be crucial factors in securing a victory for France.
\end{lstlisting}

We can observe that the model correctly picks on the context of the match.
It mentions the match in the Olympic Games between France and Croatia with a final score of 35-24 in favor of France.
It does not mention the final because, to date, we do not have a way to include such information in a reliable way yet.
Instead, it is able to understand the high importance that Olympic Games represent in handball, as the most prestigious competition.
We can also notice that the model highlights facts related to Dika Mem and Elohim Prandi who are key actors of the French strategies.
Additionally, the LLM mentions the name of the wing players for the ability to play fast breaks which aligned with the reputation of the team's dynamics.

The response finally highlights the impact of some Croatian players such as the center back Igor Karacic, who was part of the All-Star team of the 2020 European championships.
Unsurprisingly, the conclusion is that the match will be rather in favor of France and the key element of the victory will be about the ability to capitalize on the strong defense.

\section{Conclusion} \label{sec:conclusion}

In this work, we leveraged different AI tools to predict, explain and generate human-friendly explanations for the handball tournament at the 2024 Olympic Games.
The prediction model is based on a deep neural network architecture and is articulated around transfer learning to benefit from the knowledge acquired by another model pre-trained on clubs data.
Though we face challenges because of the volume of data available, this approach allows us to generate reasonable outcomes that are in line with sports experts expectations.

To further improve the performance of the prediction model, continuing the data collection to improve the quality as well as the quantity will be a key enabler.
This will imply heavy manual data collection tasks or the identification of improved data sources, though the bottleneck is often coming directly from the handball federations which do not have a systematic reporting in some competitions (in particular friendly matches).

Another path for improvement would be to develop a dedicated model that will be used to feed the embedding of the final prediction model.
In particular, if data allow, one could build a model that focuses on predicting the performance and statistics of the players during a match (e.g. goals scored and blocks).
This model's embedding will be explicitly focused on players performance and can be another way of pre-training.
As mentioned in the previous paragraph, this is dependent on the data quantity and quality made available by the different federations.

Focusing on the improvements of the AI-generated explanations, another step would be to further explore the model selection.
In this work, we leveraged a quantized version of Mistral 7B Instruct to lower the computational footprint.
Indeed, in practice, only few clubs or federations would have the required budget to afford paying a dedicated API, fine-tuning their own model or even having the necessary compute resources.
All of the experiments were made on a relatively small server\footnote{The server specs are: GPU RTX 3060 Ti with 8GB of GPU memory and 40GB of CPU RAM} for this exact consideration.
Additional compute resources would allow bigger models, more detailed instructions in the prompt template or potentially fine-tune the model on sports related datasets.

Our proposed solution leverages advanced AI technologies to generate accurate and meaningful predictions applied to handball at the 2024 Olympic Games.
Not only it allows to generate accurate predictions about the final score of a match but it also comes up with a detailed explanations that can be shared with sports experts and coaches so they can better understand how they would need to adapt their strategy for the match.

\section*{Disclaimer}

This work is not related to Amazon.

\printbibliography

@misc{jiang_mistral_2023,
	title = {Mistral {7B}},
	url = {http://arxiv.org/abs/2310.06825},
	doi = {10.48550/arXiv.2310.06825},
	publisher = {arXiv},
	author = {Jiang, Albert Q. and Sablayrolles, Alexandre and Mensch, Arthur and Bamford, Chris and Chaplot, Devendra Singh and Casas, Diego de las and Bressand, Florian and Lengyel, Gianna and Lample, Guillaume and Saulnier, Lucile and Lavaud, Lélio Renard and Lachaux, Marie-Anne and Stock, Pierre and Scao, Teven Le and Lavril, Thibaut and Wang, Thomas and Lacroix, Timothée and Sayed, William El},
	month = oct,
	year = {2023},
	note = {arXiv:2310.06825 [cs]},
}

@misc{felice_predicting_2023,
	title = {Predicting {Handball} {Matches} with {Machine} {Learning} and {Statistically} {Estimated} {Team} {Strengths}},
	copyright = {All rights reserved},
	doi = {10.48550/arXiv.2307.11777},
	publisher = {arXiv},
	author = {Felice, Florian and Ley, Christophe},
	year = {2023},
}

@article{felice_ranking_2024,
	title = {Ranking handball teams from statistical strength estimation},
	copyright = {All rights reserved},
	issn = {0943-4062, 1613-9658},
	url = {https://link.springer.com/10.1007/s00180-024-01522-0},
	doi = {10.1007/s00180-024-01522-0},
	journal = {Computational Statistics},
	author = {Felice, Florian},
	month = jun,
	year = {2024},
}

@misc{felice_statistically_2023,
	title = {Statistically {Enhanced} {Learning}: a feature engineering framework to boost (any) learning algorithms},
	copyright = {All rights reserved},
	shorttitle = {Statistically {Enhanced} {Learning}},
	doi = {10.48550/arXiv.2306.17006},
	publisher = {arXiv},
	author = {Felice, Florian and Ley, Christophe and Groll, Andreas and Bordas, Stéphane},
	year = {2023},
}

@article{cheng_snil_2024,
	title = {{SNIL}: {Generating} {Sports} {News} {From} {Insights} {With} {Large} {Language} {Models}},
	issn = {1941-0506},
	shorttitle = {{SNIL}},
	url = {https://ieeexplore.ieee.org/abstract/document/10507016},
	doi = {10.1109/TVCG.2024.3392683},
	journal = {IEEE Transactions on Visualization and Computer Graphics},
	author = {Cheng, Liqi and Deng, Dazhen and Xie, Xiao and Qiu, Rihong and Xu, Mingliang and Wu, Yingcai},
	year = {2024},
	note = {Conference Name: IEEE Transactions on Visualization and Computer Graphics},
	pages = {1--14},
}

@misc{xia_sportqa_2024,
	title = {{SportQA}: {A} {Benchmark} for {Sports} {Understanding} in {Large} {Language} {Models}},
	shorttitle = {{SportQA}},
	url = {http://arxiv.org/abs/2402.15862},
	doi = {10.48550/arXiv.2402.15862},
	publisher = {arXiv},
	author = {Xia, Haotian and Yang, Zhengbang and Wang, Yuqing and Tracy, Rhys and Zhao, Yun and Huang, Dongdong and Chen, Zezhi and Zhu, Yan and Wang, Yuan-fang and Shen, Weining},
	month = jun,
	year = {2024},
	note = {arXiv:2402.15862 [cs]},
}

@misc{noauthor_generative_nodate,
	title = {Generative {AI} - {Worldwide} {\textbar} {Statista} {Market} {Forecast}},
	url = {https://www.statista.com/outlook/tmo/artificial-intelligence/generative-ai/worldwide},
	journal = {Statista},
}

@inproceedings{tan_survey_2018,
	address = {Cham},
	title = {A {Survey} on {Deep} {Transfer} {Learning}},
	isbn = {978-3-030-01424-7},
	doi = {10.1007/978-3-030-01424-7_27},
	booktitle = {Artificial {Neural} {Networks} and {Machine} {Learning} – {ICANN} 2018},
	publisher = {Springer International Publishing},
	author = {Tan, Chuanqi and Sun, Fuchun and Kong, Tao and Zhang, Wenchang and Yang, Chao and Liu, Chunfang},
	editor = {Kůrková, Věra and Manolopoulos, Yannis and Hammer, Barbara and Iliadis, Lazaros and Maglogiannis, Ilias},
	year = {2018},
	pages = {270--279},
}

@misc{sundararajan_axiomatic_2017,
	title = {Axiomatic {Attribution} for {Deep} {Networks}},
	url = {http://arxiv.org/abs/1703.01365},
	doi = {10.48550/arXiv.1703.01365},
	publisher = {arXiv},
	author = {Sundararajan, Mukund and Taly, Ankur and Yan, Qiqi},
	month = jun,
	year = {2017},
	note = {arXiv:1703.01365 [cs]},
}

@article{zhang_benchmarking_2024,
	title = {Benchmarking {Large} {Language} {Models} for {News} {Summarization}},
	volume = {12},
	issn = {2307-387X},
	url = {https://doi.org/10.1162/tacl_a_00632},
	doi = {10.1162/tacl_a_00632},
	journal = {Transactions of the Association for Computational Linguistics},
	author = {Zhang, Tianyi and Ladhak, Faisal and Durmus, Esin and Liang, Percy and McKeown, Kathleen and Hashimoto, Tatsunori B.},
	month = jan,
	year = {2024},
	pages = {39--57},
}

@article{wang_large_2023,
	title = {Large {Language} {Models} {Are} {Latent} {Variable} {Models}: {Explaining} and {Finding} {Good} {Demonstrations} for {In}-{Context} {Learning}},
	volume = {36},
	shorttitle = {Large {Language} {Models} {Are} {Latent} {Variable} {Models}},
	url = {https://proceedings.neurips.cc/paper_files/paper/2023/hash/3255a7554605a88800f4e120b3a929e1-Abstract-Conference.html},
	journal = {Advances in Neural Information Processing Systems},
	author = {Wang, Xinyi and Zhu, Wanrong and Saxon, Michael and Steyvers, Mark and Wang, William Yang},
	month = dec,
	year = {2023},
	pages = {15614--15638},
}

@inproceedings{addlesee_grounding_2024,
	address = {Torino, Italia},
	title = {Grounding {LLMs} to {In}-prompt {Instructions}: {Reducing} {Hallucinations} {Caused} by {Static} {Pre}-training {Knowledge}},
	shorttitle = {Grounding {LLMs} to {In}-prompt {Instructions}},
	url = {https://aclanthology.org/2024.safety4convai-1.1},
	booktitle = {Proceedings of {Safety4ConvAI}: {The} {Third} {Workshop} on {Safety} for {Conversational} {AI} @ {LREC}-{COLING} 2024},
	publisher = {ELRA and ICCL},
	author = {Addlesee, Angus},
	editor = {Dinkar, Tanvi and Attanasio, Giuseppe and Curry, Amanda Cercas and Konstas, Ioannis and Hovy, Dirk and Rieser, Verena},
	month = may,
	year = {2024},
	pages = {1--7},
}

@misc{hu_sportsmetrics_2024,
	title = {{SportsMetrics}: {Blending} {Text} and {Numerical} {Data} to {Understand} {Information} {Fusion} in {LLMs}},
	shorttitle = {{SportsMetrics}},
	url = {http://arxiv.org/abs/2402.10979},
	doi = {10.48550/arXiv.2402.10979},
	publisher = {arXiv},
	author = {Hu, Yebowen and Song, Kaiqiang and Cho, Sangwoo and Wang, Xiaoyang and Foroosh, Hassan and Yu, Dong and Liu, Fei},
	month = jun,
	year = {2024},
	note = {arXiv:2402.10979 [cs]},
}

@book{goodfellow_deep_2016,
	title = {Deep {Learning}},
	isbn = {978-0-262-33737-3},
	publisher = {MIT Press},
	author = {Goodfellow, Ian and Bengio, Yoshua and Courville, Aaron},
	month = nov,
	year = {2016},
}

@misc{karlis_modelling_2024,
	title = {Modelling handball outcomes using univariate and bivariate approaches},
	url = {http://arxiv.org/abs/2404.04213},
	doi = {10.48550/arXiv.2404.04213},
	publisher = {arXiv},
	author = {Karlis, Dimitris and Michels, Rouven and Otting, Marius},
	month = apr,
	year = {2024},
	note = {arXiv:2404.04213 [stat]},
}

@misc{groll_prediction_2019,
	title = {Prediction of the 2019 {IHF} {World} {Men}'s {Handball} {Championship} - {An} underdispersed sparse count data regression model},
	url = {http://arxiv.org/abs/1901.05722},
	doi = {10.48550/arXiv.1901.05722},
	publisher = {arXiv},
	author = {Groll, Andreas and Heiner, Jonas and Schauberger, Gunther and Uhrmeister, Jörn},
	month = jan,
	year = {2019},
	note = {arXiv:1901.05722 [stat]},
}

@misc{wang_player2vec_2024,
	title = {player2vec: {A} {Language} {Modeling} {Approach} to {Understand} {Player} {Behavior} in {Games}},
	shorttitle = {player2vec},
	url = {http://arxiv.org/abs/2404.04234},
	doi = {10.48550/arXiv.2404.04234},
	publisher = {arXiv},
	author = {Wang, Tianze and Honari-Jahromi, Maryam and Katsarou, Styliani and Mikheeva, Olga and Panagiotakopoulos, Theodoros and Asadi, Sahar and Smirnov, Oleg},
	month = jun,
	year = {2024},
	note = {arXiv:2404.04234 [cs]},
}

@misc{vaswani_attention_2023,
	title = {Attention {Is} {All} {You} {Need}},
	url = {http://arxiv.org/abs/1706.03762},
	doi = {10.48550/arXiv.1706.03762},
	publisher = {arXiv},
	author = {Vaswani, Ashish and Shazeer, Noam and Parmar, Niki and Uszkoreit, Jakob and Jones, Llion and Gomez, Aidan N. and Kaiser, Lukasz and Polosukhin, Illia},
	month = aug,
	year = {2023},
	note = {arXiv:1706.03762 [cs]},
}

@article{hvattum_comprehensive_2019,
	title = {A comprehensive review of plus-minus ratings for evaluating individual players in team sports},
	volume = {18},
	copyright = {http://creativecommons.org/licenses/by-nc-nd/3.0},
	issn = {1684-4769},
	url = {https://www.sciendo.com/article/10.2478/ijcss-2019-0001},
	doi = {10.2478/ijcss-2019-0001},
	number = {1},
	journal = {International Journal of Computer Science in Sport},
	author = {Hvattum, Lars Magnus},
	month = jul,
	year = {2019},
	pages = {1--23},
}

@inproceedings{ansel_pytorch_2024,
	address = {La Jolla CA USA},
	title = {{PyTorch} 2: {Faster} {Machine} {Learning} {Through} {Dynamic} {Python} {Bytecode} {Transformation} and {Graph} {Compilation}},
	isbn = {9798400703850},
	shorttitle = {{PyTorch} 2},
	url = {https://dl.acm.org/doi/10.1145/3620665.3640366},
	doi = {10.1145/3620665.3640366},
	booktitle = {Proceedings of the 29th {ACM} {International} {Conference} on {Architectural} {Support} for {Programming} {Languages} and {Operating} {Systems}, {Volume} 2},
	publisher = {ACM},
	author = {Ansel, Jason and Yang, Edward and He, Horace and Gimelshein, Natalia and Jain, Animesh and Voznesensky, Michael and Bao, Bin and Bell, Peter and Berard, David and Burovski, Evgeni and Chauhan, Geeta and Chourdia, Anjali and Constable, Will and Desmaison, Alban and DeVito, Zachary and Ellison, Elias and Feng, Will and Gong, Jiong and Gschwind, Michael and Hirsh, Brian and Huang, Sherlock and Kalambarkar, Kshiteej and Kirsch, Laurent and Lazos, Michael and Lezcano, Mario and Liang, Yanbo and Liang, Jason and Lu, Yinghai and Luk, C. K. and Maher, Bert and Pan, Yunjie and Puhrsch, Christian and Reso, Matthias and Saroufim, Mark and Siraichi, Marcos Yukio and Suk, Helen and Zhang, Shunting and Suo, Michael and Tillet, Phil and Zhao, Xu and Wang, Eikan and Zhou, Keren and Zou, Richard and Wang, Xiaodong and Mathews, Ajit and Wen, William and Chanan, Gregory and Wu, Peng and Chintala, Soumith},
	month = apr,
	year = {2024},
	pages = {929--947},
}

@misc{frantar_gptq_2023,
	title = {{GPTQ}: {Accurate} {Post}-{Training} {Quantization} for {Generative} {Pre}-trained {Transformers}},
	shorttitle = {{GPTQ}},
	url = {http://arxiv.org/abs/2210.17323},
	doi = {10.48550/arXiv.2210.17323},
	publisher = {arXiv},
	author = {Frantar, Elias and Ashkboos, Saleh and Hoefler, Torsten and Alistarh, Dan},
	month = mar,
	year = {2023},
	note = {arXiv:2210.17323 [cs]},
}

\section{Appendix}

\subsection{Prompt template}\label{apdx:prompt}

\begin{lstlisting}[caption=Prompt template]
(*\label{code:prompt_template}*)
<s>[INST] You are a sport assistant for a handball team. You are working with the head coach who is interested in the prediction of the handball match about to happen, as well as understanding the main drivers of the score. You are a machine learning expert with experience in sports predictions.

# Inputs

## Match information
This part gives you global context about the match (teams, competition, date/time and predicted score):
{info}

## Feature descriptions
The bullet points below list the potential features used by the model and their meaning (other features are player names). For each feature, you have the explanations of what a positive attribution means (you can guess their definition for negative attributions):
{feat}

## Explanations
The below information will give you the details and explanations about the prediction.
{explain}

# Examples
Report 1: In this international friendly game, we expect to see Norway be defeated by Denmark by a small margin (31-32). Given the different strengths of both teams, we can expect to see a close match where we can expect Denmark to win but other outcomes (such as a draw) are still possible.
We report that Magnus Saugstrup (line player) and Mikkel Hansen (left back) are marked as important drivers of the score for Denmark, highlighting their individual performances as well as their strong connection to hurt the Norwegian defense.
We also note that on the Norwegian side, Christian O'Sullivan and Magnus Gullerud have a strong negative impact on Denmark's ability to score which translates their defense strengths.
Overall, during this friendly game, we expect to see a very closed game with no team really able to take the best over their opponent.
The team that will firmly defend, primarily in the central sector, while ensuring a good synergy between the back and line players can expect to end victorious.

Report 2: In this LFH Division 1, Women match, we expect to see Metz Handball win against Plan de Cuques by a pretty large margin (35-22). The attack strength of Metz Handball is one of the main drivers of the final prediction. This means that the high and stable average of scored goals by Metz might not be challenged by the defense from Plan de Cuques.
Playing a positive role for Metz Handball, we expect to see Sarah Bouktit have a strong impact on the match as a line player translating her strong presence in the opponent's defense. We also notice that wing players from Metz (Chloe Valentini and Lucie Granier) are expected to score several goals suggesting their ability to run for fast breaks and get the best of attack on their wing corners.
Benefiting Plan de Cuques' prediction, Tatiana Elisme (on the right wing) and Daphne Gautschi (center back) positively impact the prediction for Plan de Cuques, suggesting individual performances from the team are the only alternative to counteract against the leader Metz Handball.
Overall, the match seems to be one-sided where we expect to see Metz take an important advantage against their opponent and demonstrate their strong attack a limit conceded goals thanks to their powerful defense.

# Instructions
The explanations in the inputs contain the list of all variables used to predict the number of goals aligned with the attribution score. The attribution score measures the impact of each feature on the predicted number of goals. Therefore, the higher the value the higher the impact on the number of goals {team} will score. Conversely, when the value is negative, it means that the attribution of that feature has a negative impact on the ability of {team} to score. The values you observe are the feature attributions (their local importance to the prediction) and not the actual feature values. The feature set is composed of 2 main types: regular predictive features and player names.
If you identify patterns on groups of players from similar positions (back, wing, line players) in the same team impacting the match, you can highlight these facts. You can remember that line players (also called pivots) highly impact the performance of back and center back players (by creating blocks or receiving passes from them). Do not hesistate to highlight few names of players for whom the coach will need to keep a close look at.
Your task is to generate a short report of the match prediction with the main drivers impacting either positively or negatively {team}. The objective it to send it to the head coach of the team {team}, so use simple words, do not include nor refer to the values of attribution scores and follow the same structure as in the examples. Avoid bullet points/lists and instead prefer plain text. Be concise.
[/INST]
\end{lstlisting}

\subsection{Match importance values}\label{apdx:importance}

\begin{table}[!ht]
  \centering
  \begin{tabular}{ccc}
    \toprule
    Value & National teams & Clubs \\
    \midrule
    10 & Olympic Games &  \\
    9 & World championships &  \\
    8 & European championships &  \\
    7 & African cup / Eurocup &  \\
    6 & Asian cup / Qualifiers & EHF Champions League \\
    5 & Tournaments / Emerging nations & EHF European League \\
    4 & International Friendly Games & EHF European Cup \\
    3 & & Regular championships \\
    2 & & National cups \\
    1 & & Friendly games \\
    \bottomrule
  \end{tabular}
  \caption{Values of different leagues for national teams and clubs}\label{tab:importance}
\end{table}

\newpage

\subsection{Predictions from quarter finals}\label{apdx:results}

\subsubsection{Men's tournament}

\begin{figure}[!ht]
    \centering
    \input{figures/tex/PredictionsQuarterFinalsMen}
    \caption{Predictions for the final round of the 2024 Olympic Games for the men's tournament}
    \label{fig:final_round_men}
\end{figure}

\begin{table}[!ht]
    \begin{subtable}{.5\linewidth}
      \centering
        \begin{tabular}{>{\rowmac}c>{\rowmac}l>{\rowmac}c>{\rowmac}c<{\clearrow}}
            \toprule
            Rank & Team & Points & Goal difference \\
            \midrule
            \setrow{\bfseries} 1 & \worldflag[length=0.5cm, width=0.3cm]{HR} Croatia & 9 & 36 \\
            \setrow{\bfseries} 2 & \worldflag[length=0.5cm, width=0.3cm]{ES} Spain & 8 & 20 \\
            \setrow{\bfseries} 3 & \worldflag[length=0.5cm, width=0.3cm]{DE} Germany & 7 & 5 \\
            \setrow{\bfseries} 4 & \worldflag[length=0.5cm, width=0.3cm]{SE} Sweden & 3 & 7 \\
            5 & \worldflag[length=0.5cm, width=0.3cm]{SI} Slovenia & 3 & 1 \\
            6 & \worldflag[length=0.5cm, width=0.3cm]{JP} Japan & 0 & -15 \\
            \bottomrule
          \end{tabular}\label{tab:gpA_men_rank}
          \caption{Group A}
    \end{subtable}%
    \begin{subtable}{.5\linewidth}
      \centering
        \begin{tabular}{>{\rowmac}c>{\rowmac}l>{\rowmac}c>{\rowmac}c<{\clearrow}}
            \toprule
            Rank & Team & Points & Goal difference \\
            \midrule
            \setrow{\bfseries} 1 & \worldflag[length=0.5cm, width=0.3cm]{DK} Denmark & 7 & 20 \\
            \setrow{\bfseries} 2 & \worldflag[length=0.5cm, width=0.3cm]{FR} France & 7 & 16 \\
            \setrow{\bfseries} 3 & \worldflag[length=0.5cm, width=0.3cm]{NO} Norway & 6 & 2 \\
            \setrow{\bfseries} 4 & \worldflag[length=0.5cm, width=0.3cm]{HU} Hungary & 5 & 0 \\
            5 & \worldflag[length=0.5cm, width=0.3cm]{EG} Egypt & 5 & -2 \\
            6 & \worldflag[length=0.5cm, width=0.3cm]{AR} Argentina & 0 & -13 \\
            \bottomrule
        \end{tabular}\label{tab:gpB_men_rank}
        \caption{Group B}
    \end{subtable}
    \caption{Predicted ranking after preliminary round of the 2024 Olympic Games for the men's tournament. Teams in bold are qualified for the next round.}
    \label{tab:rank_prelim_men}
\end{table}

\newpage

\subsubsection{Women's tournament}

\begin{figure}[!ht]
    \centering
    \input{figures/tex/PredictionsQuarterFinalsWomen}
    \caption{Predictions for the final round of the 2024 Olympic Games for the women's tournament}
    \label{fig:final_round_women}
\end{figure}

\begin{table}[!ht]
    \begin{subtable}{.5\linewidth}
      \centering
        \begin{tabular}{>{\rowmac}c>{\rowmac}l>{\rowmac}c>{\rowmac}c<{\clearrow}}
            \toprule
            Rank & Team & Points & Goal difference \\
            \midrule
            \setrow{\bfseries} 1 & \worldflag[length=0.5cm, width=0.3cm]{NO} Norway & 8 & 37 \\
            \setrow{\bfseries} 2 & \worldflag[length=0.5cm, width=0.3cm]{SE} Sweden & 8 & 20 \\
            \setrow{\bfseries} 3 & \worldflag[length=0.5cm, width=0.3cm]{DE} Germany & 6 & 25 \\
            \setrow{\bfseries} 4 & \worldflag[length=0.5cm, width=0.3cm]{DK} Denmark & 6 & 21 \\
            5 & \worldflag[length=0.5cm, width=0.3cm]{KR} South Korea & 2 & -4 \\
            6 & \worldflag[length=0.5cm, width=0.3cm]{SI} Slovenia & 0 & -9 \\
            \bottomrule
          \end{tabular}\label{tab:gpA_women_rank}
          \caption{Group A}
    \end{subtable}%
    \begin{subtable}{.5\linewidth}
      \centering
        \begin{tabular}{>{\rowmac}c>{\rowmac}l>{\rowmac}c>{\rowmac}c<{\clearrow}}
            \toprule
            Rank & Team & Points & Goal difference \\
            \midrule
            \setrow{\bfseries} 1 & \worldflag[length=0.5cm, width=0.3cm]{ES} Spain & 9 & 11 \\
            \setrow{\bfseries} 2 & \worldflag[length=0.5cm, width=0.3cm]{FR} France & 7 & 37 \\
            \setrow{\bfseries} 3 & \worldflag[length=0.5cm, width=0.3cm]{HU} Hungary & 6 & 24 \\
            \setrow{\bfseries} 4 & \worldflag[length=0.5cm, width=0.3cm]{BR} Brazil & 5 & 10 \\
            5 & \worldflag[length=0.5cm, width=0.3cm]{NL} Netherlands & 3 & 5 \\
            6 & \worldflag[length=0.5cm, width=0.3cm]{AO} Angola & 0 & -10 \\
            \bottomrule
        \end{tabular}\label{tab:gpB_women_rank}
        \caption{Group B}
    \end{subtable}
    \caption{Predicted ranking after preliminary round of the 2024 Olympic Games for the women's tournament. Teams in bold are qualified for the next round.}
    \label{tab:rank_prelim_women}
\end{table}

\newpage

\subsubsection{Local explanation of France - Croatia (men's final)}\label{apdx:local_explain_ig}

\begin{figure}[!ht]
    \centering
    \begin{tikzpicture}
    \pgfplotstableread[col sep = comma]{figures/data/FeatImport.csv}\tabledm
    \def\xlistmacro{}
    \def\xliststring{}
    \pgfplotstableforeachcolumnelement{y}\of\tabledm\as\entry{%
    \xifinlist{\entry}{\xlistmacro}{}{
            \listxadd{\xlistmacro}{\entry}
            \edef\xliststring{\xliststring\entry,}
        }
    }
  \begin{axis}[
    colorbar, colormap name=greenred,
    colorbar style={
            ylabel style={anchor=west, text width=3cm, rotate=270},
        },
    scatter,
    scatter src=x,
    only marks,
    clip mode=individual,
    scatter/@pre marker code/.append code={
            \pgfkeys{/pgf/fpu=true,/pgf/fpu/output format=fixed}
            \pgfmathsetmacro\negheight{-\pgfplotspointmeta}         
            \fill [draw=black] 
             (axis direction cs:0,0.3) rectangle (axis direction cs:\negheight,-0.3);
            \pgfplotsset{mark=none}
        },
    width=11cm,
    height=12cm,
    legend style={at={(0.5,-0.15)},
      anchor=north,legend columns=-1},
    x tick label style={draw=none, anchor=north}, 
    y tick label style={font=\tiny, draw=none, anchor=east},
    xlabel = {Attribution},
    ytick=data,
    yticklabels from table={\tabledm}{y}
    ]
  \addplot table[x=x,y expr=\coordindex, col sep=comma] {figures/data/FeatImport.csv};
  \end{axis}
\end{tikzpicture}
    \caption{Local explanation of the final match France vs Croatia of the men's tournament. Attribution values are calculated using Integrated Gradients from the deep neural network as presented in Section~\ref{sec:model}.}
    \label{fig:feat_import_example}
\end{figure}
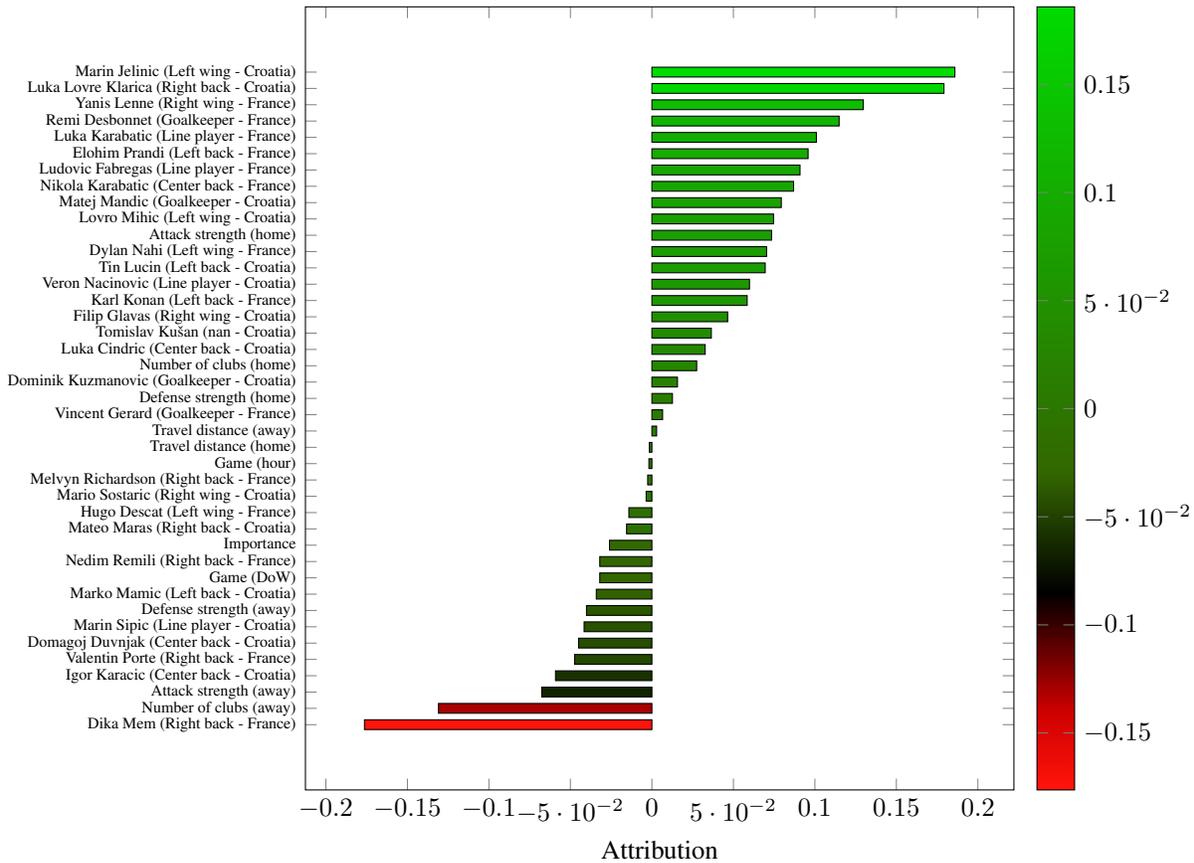

\end{document}